\documentclass[runningheads]{llncs}
\usepackage{hyperref}
\usepackage{subcaption}
\usepackage{tabularx}
\usepackage{listings}
\usepackage{verbatim}
\usepackage[table]{xcolor}
\usepackage{setspace}
\usepackage{enumitem}
\usepackage{float}
\usepackage[]{caption}
\usepackage{orcidlink}
\usepackage{pdfpages}
\usepackage{enumitem}
\begin{document}
\title{Abstractions, Scenarios, and Prompt Definitions for Process Mining with LLMs: A Case Study}
\titlerunning{Abstractions, Scenarios, and Prompts for Process Mining with LLMs}

\author{
Alessandro Berti\inst{1,2}\orcidlink{0000-0002-3279-4795},
Daniel Schuster\inst{2,1}\orcidlink{0000-0002-6512-9580},
Wil M. P. van der Aalst\inst{1,2}\orcidlink{0000-0002-0955-6940}
}
\authorrunning{A. Berti et al.}
\institute{Process and Data Science Chair, RWTH Aachen University, Aachen, Germany
\email{\{a.berti, schuster, wvdaalst\}@pads.rwth-aachen.de } \and
Fraunhofer FIT, Sankt Augustin, Germany
}

\maketitle

\begin{abstract}
Large Language Models (LLMs) are capable of answering questions in natural language for various purposes.
With recent advancements (such as GPT-4), LLMs perform at a level comparable to humans for many proficient tasks. The analysis of business processes could benefit from a natural process querying language and using the domain knowledge on which LLMs have been trained.
However, it is impossible to provide a complete database or event log as an input prompt due to size constraints.
In this paper, we apply LLMs in the context of process mining by i) abstracting the information of standard process mining artifacts
and ii) describing the prompting strategies. 
We implement the proposed abstraction techniques into \emph{pm4py}, an open-source process mining library. 
We present a case study using available event logs. Starting from different abstractions and analysis questions, we formulate prompts and evaluate the quality of the answers.
\keywords{Process Querying \and Prompting Engineering \and Large Language Models \and ChatGPT.}
\end{abstract}

\section{Introduction}
\label{sec:introduction}

\begin{figure}[ht]
\centering
\includegraphics[width=0.95\textwidth]{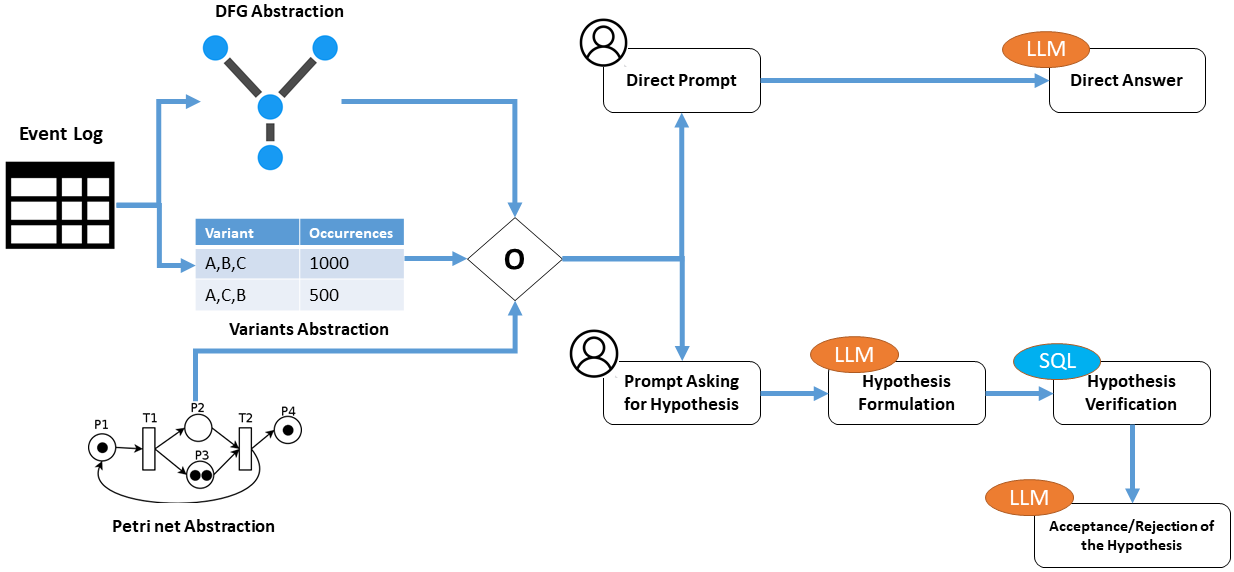}
\caption{Summary of the approach proposed in the paper: mainstream process mining artifacts can be textually abstracted and provided inside prompts directed to a LLM, such as GPT-4. Different prompting strategies can be considered.}
\label{fig:introFig}
\vspace{-6mm}
\end{figure}

Process mining uses event data from information systems to enhance business processes, involving process discovery, conformance checking, model enhancement, and predictive analytics. This data science field provides insights for improving operational processes.

Transitioning from traditional process analysis, the emergence of Large Language Models (LLMs) like GPT-4 \cite{DBLP:journals/corr/abs-2303-08774} adds a new perspective to data exploration. These advanced models, drawing on extensive training data, serve as versatile tools for general querying, enabling the extraction of valuable insights. They not only generate and retrieve information, but also hold potential to analyse and enhance business process outcomes.

In this paper, we investigate the applications of LLMs in the context of process mining, which are essential for process querying (i.e., in the verification of properties against the event log or the preprocessing phase) and in embedding the domain knowledge (used to train the LLM) in the various process mining tasks.
Despite their impressive performance, applying LLMs like GPT-4 to process mining presents challenges due to their 'context window' limitation \cite{radford2019language,brown2020language}, referring to the maximum sequence length these models can manage per interaction. 
This balancing act between information quantity and output quality can lead to significant data loss \cite{kaplan2020scaling}. Strategies including text compression, context truncation, or improved prompts \cite{reimers2019sentence,devlin2018bert} are required to effectively encapsulate process mining information.
Therefore, we explore in this paper the usage of textual abstractions of standard process mining artifacts, e.g., event logs and process models, that can embed the essential information of such artifacts.

This paper offers various prompting strategies to address the loss of information from proposed abstractions. 
A direct answer or a database query verified against the original object may be obtained, as summarized in Figure \ref{fig:introFig}. 
This study further presents the integration of the pm4py process mining library\footnote{\url{https://pm4py.fit.fraunhofer.de}} with GPT-4 and provides a case study exploring these prompting strategies using public event logs. 
The case study examines responses under different abstractions and GPT-4's domain knowledge for various processes (medical, travel expense reporting, and fines management), alongside additional process mining knowledge required for specific use cases.

The remainder of the paper is organized as follows.
Section \ref{sec:relatedWork} covers related work.
Section \ref{sec:approach} describes the abstractions and the different prompting strategies for LLMs.
Section \ref{sec:tool} describes the implementation.
Section \ref{sec:assessment} presents a case study demonstrating the usage of different abstractions and prompting strategies for process mining tasks.
Finally, Section \ref{sec:conclusion} concludes this paper.

\vspace{-3mm}
\section{Related Work}
\label{sec:relatedWork}
\vspace{-2mm}

This section provides a brief overview of process querying and the usage of domain knowledge in process mining.

Several process-mining-specific querying languages exist \cite{polyvyanyy2022process}.
In \cite{DBLP:journals/dss/PolyvyanyyOBA17}, a framework for devising process querying methods is proposed.
SQL is widely used for process discovery \cite{DBLP:conf/caise/SchonigRCJM16}, conformance checking \cite{DBLP:journals/ijais/BaaderK18}, and data preprocessing \cite{DBLP:journals/sosym/MurillasRA19}.
Cypher, a graph-based querying language, has also been adopted for process mining~\cite{DBLP:journals/jodsn/EsserF21}.
Also, Celonis PQL \cite{DBLP:books/sp/22/0001ABSGK22} is a proprietary high-performance process querying language integrated into the Celonis platform.
The mentioned languages are expressive and permit a versatile set of process mining inquiries. However, they require considerable expertise in the syntax and semantics of the query language in question and specialist knowledge.

The complexity of process querying can be reduced by translating natural language queries into database executable statements. As proposed in \cite{barbieri2022natural}, a natural language querying interface aids in democratizing process mining, making it more accessible to non-technical users. The proposed reference architecture handles natural language questions and provides responses by integrating with process mining tools, using techniques such as entity recognition and semantic parsing.
In \cite{DBLP:journals/is/KobeissiAGDBH23}, a natural language interface is proposed to assist the end-user in querying event data stored in a graph database. The natural language queries are translated into the Cypher language.
Another study, \cite{DBLP:journals/is/AaRL21}, presents a conformance checking method based on NLP, which extracts business actions and objects from textual labels associated with events. Meanwhile, \cite{DBLP:conf/otm/WinterR18} focuses on identifying constraints for business process execution from natural language documents.
In \cite{DBLP:journals/is/KechtEKR23}, chatbots are trained on customer service conversations to learn the underlying business process, showing the effectiveness of such an approach, though the generalization capabilities remain unclear.

Domain knowledge about a process can be expressed in natural language. For example, documents might contain the process execution rules if a formal model is not defined.
Utilizing domain knowledge in process discovery has been investigated in \cite{DBLP:journals/cii/SchusterZA22}.
In \cite{DBLP:conf/simpda/DixitBAHB15}, the domain knowledge of the process analyst is used to modify/improve a discovered process model.
In \cite{DBLP:journals/is/0001MW14}, an event log is abstracted to a level needed by the analyst using domain knowledge extracted from the documentation of the process to match semi-automatically
events and activities.

The role of LLMs in the business process management field has been initially investigated in \cite{DBLP:conf/bpmds/BuschRSL23}, where prompt engineering techniques to embed the required information about the business processes are discussed as an alternative to training a company/process-specific LLM.

This paper proposes the usage of LLMs for process mining tasks.
LLMs such as GPT-4 know the domain knowledge and execution constraints for the set of business processes covered by the training data.
Therefore, LLMs are not process-specific and can interpret and execute queries in natural language. In our case study, we show that the queries can be either executed directly against an abstraction of a given process mining artifact or database (SQL) queries can be automatically generated by GPT-4 to verify hypotheses.

\vspace{-3mm}
\section{Approach}
\label{sec:approach}
\vspace{-2mm}

When using LLMs for process mining, the original event logs or process model representations cannot be directly used due to size limitations.
An abstraction of these artifacts must be obtained to execute specific queries, i.e., prompts, against an LLM. 
In the following subsections, we will explain textual abstractions (see Section \ref{sec:abstraction}) and different prompt generation strategies (see Section \ref{sec:prompt_generation}).

\subsection{Abstracting Process Mining Objects}
\label{sec:abstraction}

This section describes how textual abstractions of common process mining objects, i.e., event logs and process models, can be obtained. These abstractions are later used in the proposed case study.

\lstset{
  basicstyle=\scriptsize,
  columns=fullflexible,
  frame=single,
  breaklines=true,
}

\subsubsection{Abstracting Event Logs}

Traditional event logs link each event with a single case identifier, enabling the computation of the directly-follows graph and the identification of traces and process variants \cite{DBLP:books/sp/22/WeerdtW22}. These concepts can be associated with frequency and performance metrics
\begin{itemize}[topsep=0pt]
\item In a directly-follows graph, frequency is quantified by the instances where a pair of activities are sequential, and performance is calculated as an aggregation, such as average or median, of recorded times between the two activities.
\item For a process variant, frequency is determined by the count of cases following the given trace, while performance is an aggregation, such as average or median, of total throughput times for the cases.
\end{itemize}

This information can be textually represented to aid an LLM in responding to inquiries about event data. 
Section \ref{subsec:assessHypothesis} and Listing \ref{lst:listingDfgAbstraction} demonstrate the textual representation of variants and the top 5 relationships from a Directly-Follows Graph (DFG), respectively. When constructing the directly-follows graph, various notations may be employed such as $\rightarrow$ or the phrase ``is followed by''. Despite the differences in representation, Large Language Models (LLMs) like GPT-4 interpret these notations equivalently.

\lstset{caption={Textual abstraction of a DFG.}}
\lstset{label={lst:listingDfgAbstraction}}
\begin{lstlisting}
Create Fine -> Send Fine ( frequency = 103392  performance = 7568635.65 )  
Send Fine -> Insert Fine Notification ( frequency = 79757  performance = 1501626.95 )  
Insert Fine Notification -> Add penalty ( frequency = 72334  performance = 5184000.0 )  
Add penalty -> Send for Credit Collection ( frequency = 57182  performance = 45566346.44 )  
Create Fine -> Payment ( frequency = 46952  performance = 905663.45 )
\end{lstlisting}

In the realm of object-centric event logs, wherein an event may associate with various object types, additional process modeling notations exist that can undergo textual abstraction. Specifically, object-centric directly-follows graphs \cite{DBLP:journals/sttt/BertiA23} represent an assembly of directly follows graphs corresponding to distinct object types.

\subsubsection{Abstractions of Process Models}

\begin{figure}[tb]
\centering
\includegraphics[width=\textwidth]{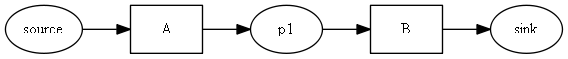}
\caption{Example sequential Petri net. From the initial state (source), the transitions A and B could be used to reach the final state (sink).}
\label{fig:sequential}
\vspace{-6mm}
\end{figure}

Formal process models, e.g., Petri nets, BPMN, and declarative models, express constraints on the activities and the paths that are executable in a process. 
For example, the Petri net shown in Fig. \ref{fig:sequential} can be abstracted as in Listing \ref{lst:listingPetriAbstraction}. The method used for textually abstracting a Petri net is not fixed and can be approached in multiple ways, provided that the naming for places and transitions is unique. The choice of abstraction strategy is arbitrary and can be tailored to specific use cases or data structures.
Similar textual abstractions of many other model formalisms (e.g., process trees, prefix trees, transition systems, BPMN models)
are possible, but we do not describe them here.

\subsection{Prompt Generation}
\label{sec:prompt_generation}

After obtaining the abstractions above, we can provide them to an LLM along with a query. 
These prompts could lead to two different types of answers, i.e.,
directly answering the original questions or leading to the formulation of hypotheses that can be verified against the data by means of database queries.

\lstset{caption={Textual abstraction of the Petri net represented in Fig. \ref{fig:sequential}.}}
\lstset{label={lst:listingPetriAbstraction}}
\begin{lstlisting}
places: [ p1, sink, source ]  
transitions: [ (A, 'A'), (B, 'B') ]  
arcs: [ (A, 'A')->p1, (B, 'B')->sink, p1->(B, 'B'), source->(A, 'A') ]  
initial marking: ['source:1']  
final marking: ['sink:1'] 
\end{lstlisting}

\subsubsection{Direct Answering}

An LLM prompt can be formulated using abstractions, such as "Describe the meaning of the activity A," which is particularly useful for descriptive or conformance checking purposes. It's important that these prompts consider no more knowledge than the provided event log or process model abstraction.

Due to the inherently probabilistic behavior of LLMs like GPT-4, the same question might yield varying responses across different sessions. This feature, rather than being an issue, is part of the model's design to promote diverse outputs and creative problem solving. If initial responses do not adequately meet the user's need, refining the question or asking more specific follow-up questions is possible to address any perceived gaps in the information provided.

\subsubsection{Hypothesis Formulation and Verification}

Certain process mining questions can be answered using the DFG/variants abstraction as they concern the order of activities. However, questions related to time and data perspectives of the event log, which require access to additional attributes or information, cannot be directly addressed by such abstractions. We may formulate hypotheses, such as impacts of specific activities on case duration, but these need further verification.

To verify a hypothesis, we can prompt an LLM, like GPT-4, with good SQL knowledge \cite{DBLP:journals/corr/abs-2303-13547}, to generate a database query that can be applied to the whole event log. The prompt uses the DFG/variants abstraction and an abstraction of event log attributes. Upon receiving the result of a query from the user, the LLM can then assess this information to confirm, refine, or dismiss the hypothesis.

It is also important to note that LLMs, provided with the top variants and attributes, can autonomously generate hypotheses on the data. Through provided abstractions, LLMs can make assertions and formulate database queries for hypothesis testing, demonstrating their flexibility and adaptability in process mining tasks.

Therefore, LLMs offer flexibility in formulating queries for hypothesis testing based on provided abstractions.

\vspace{-3mm}
\section{Implementation}
\label{sec:tool}
\vspace{-2mm}

In this section, we present the implementation of various abstractions (see Section \ref{sec:abstraction}) into the open-source process mining library, pm4py (version 2.7.5 or later). The goal is to create textual abstractions of process mining artifacts, like traditional/object-centric event logs and process models (Petri nets), suitable for GPT-4's input limit. From these abstractions, specific queries are formulated for GPT-4 execution. Listing \ref{lst:exampleUsageTraditionalObjects} demonstrates this integration, where an event log is ingested for root cause analysis, and the inductive miner algorithm discovers a process model for optimization suggestions.

\lstset{language=Python}
\lstset{frame=lines}
\lstset{caption={Example usage of the pm4py's OpenAI/GPT-4 integration on traditional process mining objects}}
\lstset{label={lst:exampleUsageTraditionalObjects}}
\lstset{basicstyle=\scriptsize}
\begin{lstlisting}
import pm4py

log = pm4py.read_xes("tests/input_data/roadtraffic100traces.xes")

iq1 = """\n What are the root causes of the performance issues in the process?
Please provide only process and data specific considerations,
no general considerations."""
print(pm4py.llm.abstract_variants(log) + iq1)

net, im, fm = pm4py.discover_petri_net_inductive(log)


iq2 = """\n Can you provide suggestions to improve the process model
based on your domain knowledge?"""
print(pm4py.llm.abstract_petri_net(net, im, fm) + iq2)
\end{lstlisting}

\vspace{-3mm}
\section{Case Study}
\label{sec:assessment}
\vspace{-2mm}

We present a case study using publicly available event logs and GPT-4~\cite{DBLP:journals/corr/abs-2303-08774}.
We propose an assessment of prompts that can be directly answered by GPT-4.
Further, we propose an example of hypothesis formulation and verification against the entire dataset (by means of a SQL query).

\vspace{-2mm}
\subsection{Direct Answering}
\label{susec:directAnswering}

To assess prompts requiring direct answers from the LLM, we use publicly available event logs: (1) Road Traffic Fine Management process \footnote{\url{https://data.4tu.nl/articles/_/12683249/1}}, which is related to the management of fines in an Italian municipality, (2) BPI Challenge 2020: Domestic Declarations \footnote{\url{https://data.4tu.nl/collections/_/5065541/1}}, which is a travel expense approval process, (3) Sepsis Cases\footnote{\url{https://data.4tu.nl/articles/_/12707639/1}}, which is a medical process for sepsis treatment, and (4) Conformance Checking Challenge 2019\footnote{\url{https://data.4tu.nl/articles/_/12707639/1}}, which is a medical training process.

\begin{table}[tb]
\caption{Experimental results for the provided prompts (each containing an abstraction and a question) on publicly available event logs.}
\centering
\scriptsize
\begin{tabular}{|l|c|c|c|c|c|}
\hline
{\bf Question} & {\bf Abstraction} & {\bf Road Traffic} & {\bf BPIC 2020} & {\bf Sepsis} & {\bf CCC 19} \\
\hline
\multicolumn{6}{|l|}{\bfseries Descriptive Questions} \\
\hline
DQ1 & DFG & \cellcolor{green}~ & \cellcolor{green}~ & \cellcolor{green}~ & \cellcolor{green}~ \\
DQ1 & Variants & \cellcolor{green}~ & \cellcolor{green}~ & \cellcolor{green}~ & \cellcolor{green}~ \\
\hline
\multicolumn{6}{|l|}{\bfseries Conformance Questions} \\
\hline
CQ1 & DFG & \cellcolor{green}~ & \cellcolor{green}~ & \cellcolor{orange}~ & \cellcolor{red}~ \\
CQ1 & Variants & \cellcolor{green}~ & \cellcolor{green}~ & \cellcolor{red}~ & \cellcolor{orange}~ \\
\hline
\multicolumn{6}{|l|}{\bfseries Process Improvement Questions} \\
\hline
IQ1 & DFG & \cellcolor{orange}~ & \cellcolor{orange}~ & \cellcolor{orange}~ & \cellcolor{orange}~ \\
IQ1 & Variants & \cellcolor{green}~ & \cellcolor{green}~ & \cellcolor{red}~ & \cellcolor{red}~ \\
\hline
IQ2 & Petri net & \cellcolor{orange}~ & \cellcolor{green}~ & \cellcolor{green}~ & \cellcolor{red}~ \\
\hline
\end{tabular}
\label{tab:experimentalResults}
\vspace{-4mm}
\end{table}

We have compiled a list of questions related to processes, sorted into various categories\footnote{A more extensive list of questions is available at \url{https://pm4py.fit.fraunhofer.de/static/assets/api/2.7.3/api.html\#openai-integration-pm4py-openai}.}. Each question is accompanied by acceptance criteria to help determine if the response given by the LLM is satisfactory.

\noindent\textbf{Descriptive Questions:}
\begin{enumerate}[label={\bf DQ\arabic*},leftmargin=2\parindent]
\item Can you describe the process contained in this data?
\begin{itemize}
    \item GPT-4 should provide the name/category of the process underlying the data and the description of the main steps of the process).
    \item If GPT-4 does not correctly understand the context and identifies the wrong name or category for the process, the response is considered unsatisfactory.
\end{itemize}
\end{enumerate}

\noindent\textbf{Conformance Questions:}
\begin{enumerate}[label={\bf CQ\arabic*},leftmargin=2\parindent]
\item Can you pinpoint the central anomalies of the process from this data? Please only process and data-specific considerations, not general considerations.
\begin{itemize}
    \item Our expectation is that GPT-4, using its domain knowledge,
is able to identify paths that are illogical, rework, or missing activities.
    \item A response is deemed unsatisfactory if GPT-4 points to infrequent activities/paths, and to paths with high performance, without exploiting the domain knowledge about the process.
\end{itemize}
\end{enumerate}

\noindent\textbf{Process Improvement Questions:}
\begin{enumerate}[label={\bf IQ\arabic*},leftmargin=2\parindent]
\item What are the root causes of performance issues specific to the process and related data? Please refrain from providing general considerations and focus on issues directly tied to the process and its data.
\begin{itemize}
    \item Our expectation is that GPT-4 should identify activities, paths, or rework
that lead to higher throughput times.
    \item A response is deemed unsatisfactory when GPT-4 identifies just the infrequent activities or paths, or is able to detect different execution orders for the activities but asks the user to verify if there is something wrong.
\end{itemize}
\item Please suggest improving the process model based on your domain knowledge. Also, please compare it against implementations of similar processes.
Provide only process and data-specific considerations, not general ones.
\begin{itemize}
    \item We expect that GPT-4 can suggest additional activities to optimize the throughput time and reduce rework. Also, it should
be able to detect when the activities are executed in a suboptimal order.
\item A response is deemed unsatisfactory if general considerations about merging activities or reducing invisible steps are provided.
\end{itemize}
\end{enumerate}

Certain queries align closely with those presented in \cite{barbieri2022natural}. Specifically, IQ1 and IQ2 correspond to questions 104 and 71 respectively, as listed in the provided resource (\url{https://ic.unicamp.br/~luciana.barbieri/pmquestions.csv}). Nevertheless, DQ1 and CQ1, which pertain to descriptive analytics and anomaly detection, exceed the capabilities offered by the Everflow tool.

All the considered prompts have been created starting from the result of abstraction and including one question. 
The prompts have been executed against GPT-4 \cite{DBLP:journals/corr/abs-2303-08774}.
We report on the quality of the answers in Table \ref{tab:experimentalResults}. 
Each row reports the results on the considered logs for a given abstraction and question.
Different colors have been used: green indicates a useful answer for the given question, orange indicates a partly useful response, and red indicates that GPT-4 did not provide a good answer to the given question.

The response to descriptive questions has been satisfying for all event logs. GPT-4 for the considered questions and event logs is able to provide meaningful answers starting from both the directly-follows graph and variants. In particular, it is surprising that the DFG abstraction (which can be efficiently computed for large event logs) is sufficient not only to make local considerations (related to a path/activity of the process) but on the end-to-end execution of the process.

The conformance questions elicited satisfying responses for the "Road Traffic" and "BPI Challenge 2020" event logs (see Listing \ref{lst:anomaliesBpic2020}). Key issues like inconsistent activity order and multiple payment recordings were pinpointed for "Road Traffic," while the "BPI Challenge 2020" rework in rejections/resubmissions were identified as problematic. Contrarily, responses for medical event logs "Sepsis" and "CCC19" were below expectations, possibly due to GPT-4's limited domain knowledge of medical processes and difficulties handling their unstructured nature and high event count per case.

Responses to process improvement questions met expectations for the "Road Traffic" and "BPI Challenge 2020" event logs, especially with the variants abstraction. The directly-follows graph abstraction had limited utility in root cause analysis, as rework was more discernible in variants abstraction. However, for IQ2 employing model-based abstraction, GPT-4 offered valuable process optimization insights expressed in natural text.

For the contexts where both the DFG and variants abstraction are useful, we found that different viewpoints are provided. Therefore, the tips obtained thanks to different abstractions can be combined.
Across different executions, we obtained slightly different answers (with similar degrees of usefulness), which can lead to a broader set of insights.

\lstset{caption={Response to CQ1 for the BPI Challenge 2020 event log.}}
\lstset{label={lst:anomaliesBpic2020}}
\lstset{basicstyle=\scriptsize}
\begin{lstlisting}
From the given data, here are some of the potential anomalies I can identify in this process:

Inefficient Process Variants: There are several process variants where a declaration is rejected multiple times (by either ADMINISTRATION, PRE_APPROVER, SUPERVISOR, or BUDGET OWNER) before it is finally approved. This suggests an inefficiency in the process that may require further investigation. It is possible that the declaration is not adequately prepared before submission or the criteria for approval are not clearly communicated. These inefficient process variants not only increase the frequency of steps in the process but also increase the overall performance time.

High Performance Time: Some process variants have remarkably high performance times. For instance, the process variant where the declaration is "SUBMITTED by EMPLOYEE -> Declaration REJECTED by ADMINISTRATION" has a performance time of 13805149.2. This high performance time could indicate a delay or bottleneck in the process that needs to be addressed.

Missing Steps: There is a process variant where a declaration is directly "FINAL_APPROVED by SUPERVISOR" without going through "APPROVED by ADMINISTRATION" or "APPROVED by BUDGET OWNER" or "APPROVED by PRE_APPROVER". This could indicate a deviation from the standard process, which might either suggest an alternative process route or a process violation.

Unfinished Process: There are some process variants that do not end with "Payment Handled", such as "Declaration SUBMITTED by EMPLOYEE -> Declaration REJECTED by ADMINISTRATION -> Declaration REJECTED by EMPLOYEE" and "Declaration SUBMITTED by EMPLOYEE -> Declaration REJECTED by SUPERVISOR -> Declaration REJECTED by EMPLOYEE". This could mean that these processes are incomplete, or that the payment handling is being done outside of the tracked process.

Unexpected REJECTION: There are a few process variants where after being "FINAL_APPROVED by SUPERVISOR", the declaration is "REJECTED by MISSING". This is unexpected as the declaration should ideally not be rejected after final approval. This anomaly might indicate some errors or issues in the process or system.
\end{lstlisting}

\vspace{-3mm}
\subsection{Hypothesis Formulation and Verification}
\label{subsec:assessHypothesis}
\vspace{-2mm}

We provide an example of hypothesis formulation and verification on top of the Road Traffic Fine Management event log. For this, we formulate the prompt in Listing \ref{lst:listingHypothesisFormulation}, containing the top variants of the event log,
and a summary of the numerical attributes of the event log.

In preliminary tests, GPT-4 required details about case identifier, activity, and timestamp attributes. It also attempted to access a non-existent variant attribute and needed guidance to compute the case's duration. Yet, given the limited information, GPT-4 surprisingly formulated plausible and testable hypotheses.

GPT-4 generates various hypotheses for the given event log, including a supposed influence of the 'expense' attribute on 'Payment' activity occurrence. Testing this hypothesis using the SQL query in Listing \ref{lst:sqlCodeHypothesis} shows it to be inaccurate, as the minor difference in average expenses between cases with and without payment isn't statistically significant. Given these results, GPT-4 suggests examining the 'amount' attribute's influence on payment presence, recognizing its initial hypothesis as unsubstantiated.

\lstset{caption={Prompt provided to GPT-4 for hypothesis formulation on the Road Traffic Fine Management event log.}}
\lstset{label={lst:listingHypothesisFormulation}}
\begin{lstlisting}
If I have a process with the following process variants:
 Create Fine -> Send Fine -> Insert Fine Notification -> Add penalty -> Send for Credit Collection ( frequency = 56482  performance = 59591524.946000494 )
 Create Fine -> Payment ( frequency = 46371  performance = 889688.4000776347 )
 Create Fine -> Send Fine ( frequency = 20385  performance = 8380516.026490066 )
...

and the log of the process contains the following attributes:
amount  empty: 331240  quantiles: {0.0: 0.0, 0.25: 33.6, 0.5: 38.0, 0.75: 71.5, 1.0: 8000.0}
article  empty: 411100  quantiles: {0.0: 7.0, 0.25: 7.0, 0.5: 157.0, 0.75: 157.0, 1.0: 401.0}
expense  empty: 457483  quantiles: {0.0: 0.0, 0.25: 10.0, 0.5: 11.88, 0.75: 13.5, 1.0: 76.0}
time:timestamp  empty: 0  quantiles: {0.0: Timestamp('2000-01-01 00:00:00+0000', tz='UTC'), 0.25: Timestamp('2002-11-12 00:00:00+0000', tz='UTC'), 0.5: Timestamp('2006-02-28 00:00:00+0000', tz='UTC'), 0.75: Timestamp('2009-03-30 00:00:00+0000', tz='UTC'), 1.0: Timestamp('2013-06-18 00:00:00+0000', tz='UTC')}
...

Can you provide some hypothesis between the execution of the process and its attributes? I mean, can you provide me a DuckDB SQL query that I can execute, and return the results to you, in order for you to evaluate such hypothesis about the process? More in detail, the data is stored in a Pandas dataframe where each row is an event having the provided attributes (so there are no separate table containing the variant). Can you tell me in advance which hypothesis you want to verify? Please consider the following information: the case identifier is called "case:concept:name", the activity is stored inside the attribute "concept:name", the timestamp is stored inside the attribute "time:timestamp", the resource is stored inside the attribute "org:resource", there is not a variant column but that can be obtained as concatenation of the activities of a case, there is not a duration column but that can be obtained as difference between the timestamp of the first and the last event. Also, the dataframe is called "dataframe". You should use the EPOCH function of DuckDB to get the timestamp from the date.
\end{lstlisting}

\lstset{language=Python}
\lstset{frame=lines}
\lstset{caption={SQL code provided by GPT-4 for the verification of an hypothesis on the entire dataset}}
\lstset{label={lst:sqlCodeHypothesis}}
\lstset{basicstyle=\scriptsize}
\begin{lstlisting}
WITH case_has_payment AS (
    SELECT "case:concept:name", 
           MAX(CASE WHEN "concept:name" = 'Payment' THEN 1 ELSE 0 END)
		   as has_payment
    FROM dataframe
    GROUP BY "case:concept:name"
)
SELECT has_payment, AVG(expense) as avg_expense
FROM dataframe
JOIN case_has_payment ON
dataframe."case:concept:name" = case_has_payment."case:concept:name"
GROUP BY has_payment;
\end{lstlisting}

\vspace{-3mm}
\subsection{Limitations, Open Challenges, and Opportunities}

The results indicate that GPT-4's proficiency in addressing advanced conformance and process improvement queries improves with mainstream and standardized processes. Generally, GPT-4 exhibits substantial process mining understanding, albeit with the need for simple instructions for computing variants and throughput time. Notably, it was intriguing that GPT-4 could decipher the entire process execution from the DFG abstraction.

Nonetheless, these insights warrant validation against a wider array of questions and event logs. Additionally, the assessment of the proposed questions was based on the stated acceptance criteria, which is somewhat subjective, and alternative criteria could be employed. Consequently, the presented case study should be regarded as a preliminary exploration of LLMs' applicability in process mining.

\vspace{-3mm}
\section{Conclusion}
\label{sec:conclusion}
\vspace{-2mm}

The findings of this study provide promising indications for the application of Large Language Models (LLMs) in process mining, underscoring their potential in handling complex queries and process interpretations. LLMs, such as GPT-4, demonstrate impressive proficiency in understanding and analyzing process structures, highlighting the vast opportunities these models could bring to the field.

However, several challenges persist. One key concern is privacy - a considerable number of companies may be reticent to upload their core data to public LLMs like GPT-4 due to the sensitivity of the information involved. This brings to the fore the need for private LLMs, which can balance the utility of large-scale language models with the security needs of individual organizations.

To address privacy concerns, proprietary LLMs could be developed, trained on a mix of general and company-specific data. While current open-source models lag behind GPT-4, they're improving, suggesting the feasibility of private, customized LLMs. These models could potentially enhance process mining's efficiency and adaptability.

\vspace{-3mm}

\bibliographystyle{splncs04}
\bibliography{references}

\end{document}